# In vivo labeling and quantitative imaging of neurons using MRI

Shana Li[1], Xiang Xu[2], Canjun Li[3], Ziyan Xu[3], Qiong Ye[4], Yan Zhang[2], Chunlei Cang[3], Jie Wen[1,*]

1. Department of Radiology, The First Affiliated Hospital of USTC (Anhui Provincial Hospital), Division of Life Sciences and Medicine, University of Science and Technology of China, Hefei, Anhui, China

2. Stroke Center & Department of Neurology, The First Affiliated Hospital of USTC (Anhui Provincial Hospital), Division of Life Sciences and Medicine, University of Science and Technology of China, Hefei, Anhui, China

3. School of Life Sciences, Division of Life Sciences and Medicine, University of Science and Technology of China, Hefei, Anhui, China

4. High Magnetic Field Laboratory, Hefei Institutes of Physical Science, Chinese Academy of Sciences, Hefei, Anhui, China

* Corresponding author: Jie Wen, jiewen@ustc.edu.cn

**Summary**

Mammalian brain is a complex organ that contains billions of neurons. These neurons form various neural circuits that control the perception, cognition, emotion and behavior. Developing in vivo neuronal labeling and imaging techniques is crucial for studying the structure and function of neural circuits. In vivo techniques can provide true physiological information that cannot be provided by ex vivo methods. In this study, we describe a new strategy for in vivo neuronal labeling and quantification using MRI. To demonstrate the ability of this new method, we used neurotropic virus to deliver oatp1a1 gene to the target neural circuit. OATP1A1 protein is expressed on the neuronal membrane and can increase the uptake of a specific MRI contrast agent (Gd-EOB-DTPA). By using T1-weighted images for observation, labeled neurons "light up" on MRI. We further use a dynamic-contrast-enhancement based method to obtain measures that provide quantitative information of labeled neurons in vivo.

**Introduction**

The appeal of in vivo imaging in neuroscience is easy to understand. As one of the most complex organs, the brain contains tens of billions of neurons. These neurons are connected to each other through a special architecture to form various neural circuits (Luo, 2021) and perform different specific functions (Chen and Hong, 2018; Williams, 2016). Damaged or altered neural circuits affect brain functions and lead to various neurological diseases (Buie et al., 2010; Canter et al., 2016; Dubois et al., 2014; Hare and Duman, 2020; Hildebrand et al., 2017; Scholl et al., 2013). The brain is also a highly dynamic system and dramatic variations could happen due to functions and diseases induced changes. In vivo imaging allows the study of both structure and function of neural circuits in reasonably intact preparations, and provides a unique opportunity to monitor dynamic changes of the central nervous system.

Labeling and imaging neurons in vivo is crucial for structural and functional studies of neural circuits. Various methods have been proposed to label neurons. One of the most important methods among them relies on fluorescent protein and optical imaging techniques. Due to the intrinsic observation depth limits of optical techniques, in most cases these approaches require sacrifice of the animal and histological processing of the ex vivo tissue. As a result, a large sample size is usually required, and the brain slicing process might also cause damage of the brain tissue. Besides, in order to achieve whole brain neuronal mapping, special techniques, such as transparent brain technology (Cai et al., 2019; Chung et al., 2013), are usually required.

In this study, we describe a new strategy for in vivo neuronal labeling and quantification using MRI. To demonstrate the capability of this new method, we constructed the gene of oatp1a1 (a protein from the SLCO superfamily) onto the vector of hypotoxicity recombinant adeno-associated virus 2 retro (rAAV2-retro) (Buie et al., 2010; Tervo et al., 2016), which was subsequently used to infect the targeted neural circuit. OATP1A1 is known to be capable of transporting various small molecules across the plasma membrane, including the clinically approved hepatobillary MRI contrast agent (CA) – gadolinium-ethoxybenzyl-DTPA (Gd-EOB-DTPA) (Leonhardt et al., 2010; Mohajer et al., 2012; Patrick et al., 2014). As a result, neurons located in the targeted neural circuits are infected by rAAV2-retro and express OATP1A1 on the plasma membrane. Subsequent intrathecal administration of Gd-EOB-DTPA allows the agent to diffuse across the mouse brain along with the cerebrospinal fluid (CSF) flow. OATP1A1 proteins expressed in targeted neurons allow transmembrane uptake of Gd-EOB-DTPA. After waiting adequate time for Gd-EOB-DTPA to wash out of the brain parenchyma, labeled neurons "light up" on T1-weighted (T1W) MRI images, which provides in vivo detection of neuroanatomical connections in the mouse brain. We further recruited a dynamic contrast enhancement (DCE) based method to provide quantitative information of labeled neurons in the whole brain.

## Results

### SSp/SSs project to PO verified by ex vivo fluorescence imaging

Two viruses are used, one is the experimental group virus (rAAV2-retro-EF1α-

oatp1a1-P2A-EGFP), which expresses both oatp1a1 and EGFP in infected neurons. The other is the control group virus (rAAV2-retro-EF1α-EGFP), which only expresses EGFP. Vector structures of two viruses are shown in Figure 1A, and the recombinant viral genome plasmid map is shown in Figure S1. Viruses were injected into the posterior thalamic nuclear group (PO) area, as shown in Figure 1B. PO is an important thalamic nucleus for somatosensory information processing. Both the primary somatosensory cortex (SSp) and the secondary somatosensory cortex (SSs) areas in the cortex have neurons that project to the PO area. Twenty-one days after virus injection, brain slices were examined with fluorescence imaging. After jigsaw shooting in confocal, similar levels were shown in Figure 1C. Fluorescence results from both groups show that PO region accepts projections from SSp and SSs regions in the cortex (Figure 1C), which is consistent with previous reports (Huo et al., 2020; Oh et al., 2014).

Through the fluorescence images, we can see that OATP1A1 proteins are expressed and co-localized with EGFP in neurons (Figure 1D). Red is the immunofluorescence staining of OATP1A1 protein, green is EGFP expressed in neurons, and blue is the fluorescence staining of DAPI on the nucleus. All pictures are taken by confocal. Co-localization of the three fluorescence channels are well displayed in Figure 1D. The OATP1A1 protein is a cell membrane protein. We constructed a fusion protein to determine the location of this protein in neuronal cells. Constructed plasmid is shown in Figure 1E. Detailed plasmid map is shown in Figure S1. The recombinant plasmid was transferred into SH-SY5Y cells, and the oatp1a1-mCherry fusion protein was formed after gene expression (Figure 1F). The localization of

OATP1A1 protein within neuronal cells was examined using a fluorescence microscope. Results of fluorescence imaging and differential interference imaging support the conclusion that OATP1A1 proteins are located on the cell membrane of neuronal cells (Figure 1F). White arrows in Figure 1F marked the location of mCherry after transfection of the recombinant plasmid, indicating that OATP1A1 proteins are mainly distributed on the cell membrane of SH-SY5Y cells.

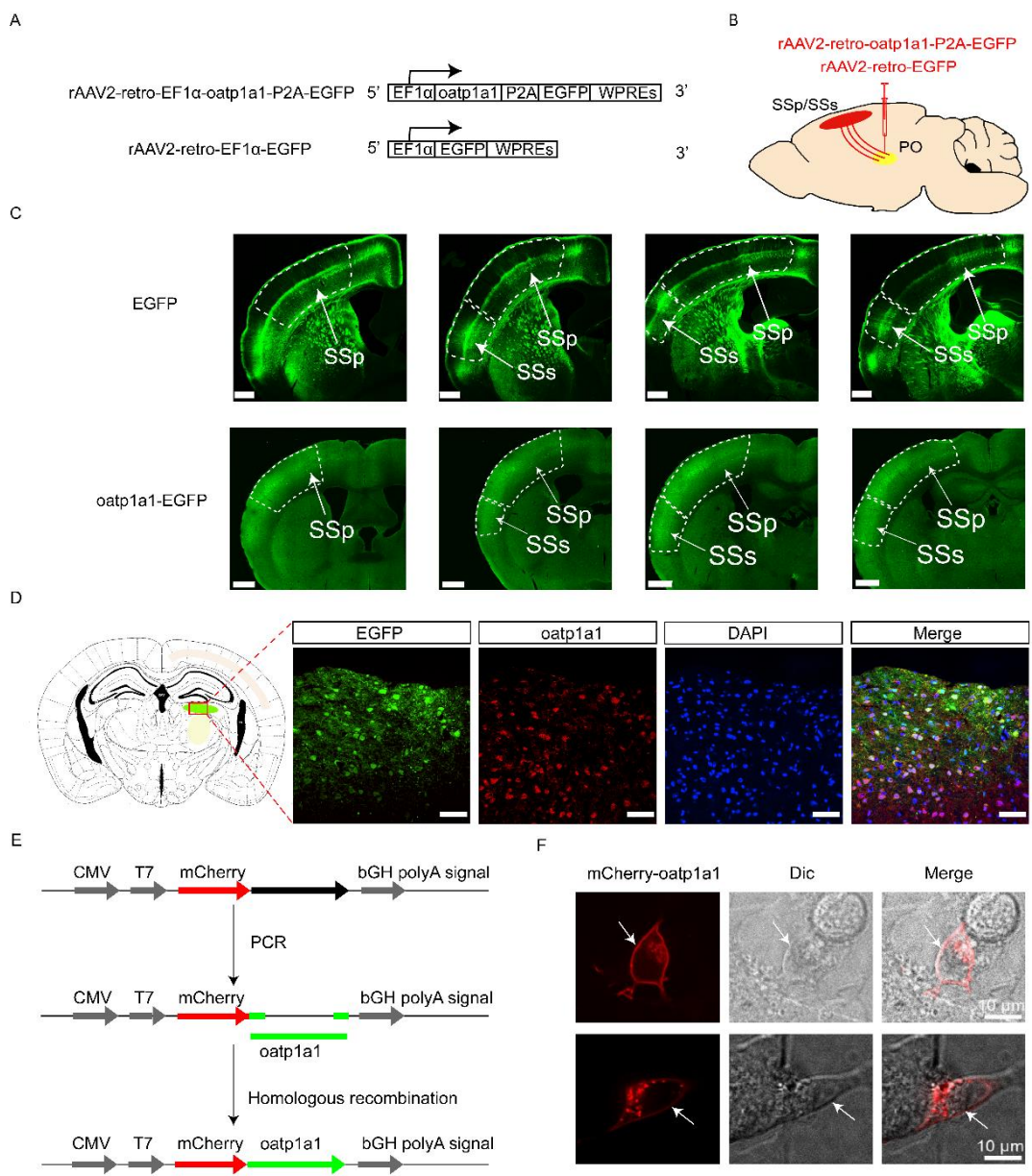

Figure 1 Ex vivo verification of neural tracing of a known neural circuit (SSp/SSs →

PO) and visualization of OATP1A1 expression on the plasma membrane. (A) Illustration of recombinant vectors of rAAV2-retro-EF1α-oatp1a1-P2A-EGFP (up) and rAAV2-retro-EF1α-EGFP (bottom). (B) Illustration of the neural circuit used for ex vivo verification. When rAAV2-retro is injected into PO, the virus will travel back to infect SSp and SSs. (C) Fluorescence images of brain slices infected with two viruses. White arrows indicate SSp and SSs areas that contain neurons projected to PO. The brain slices labeled with EGFP (up) and oatp1a1-EGFP (bottom) are taken from similar levels. Scale bars in the picture represent 500um. (D) Detection of OATP1A1 expression and co-localization with EGFP in brain tissues by immunofluorescence staining in brain tissue sections. (green: EGFP; red: OATP1A1; blue: DNA). Scale bars in the picture represent 50um. (E) Schematic diagram of plasmid construction of oatp1a1 and mCherry fusion protein. (F) Imaging of mCherry shows the localization of OATP1A1 in SH-SY5Y cells after transfection. White arrows indicate OATP1A1 proteins are mainly distributed on the cell membrane in SH-SY5Y cells. OATP1A1: Mus musculus solute carrier organic anion transporter family, member 1a1; PO: Posterior thalamic nuclear group; SSp: primary somatosensory; SSs: supplemental somatosensory.

## Brain parenchyma absorb and clear Gd-EOB-DTPA within two weeks

In order to evaluate the diffusion and metabolism of the CA into brain parenchyma, we first injected Gd-EOB-DTPA into the spinal canal of normal mice and the diffusion and metabolism of the injected CA are shown on T1W images acquired using a 14T scanner (Figure 2A). After intrathecal injection of a volume of 5-7ul, the CA concentrates at the skull base and slowly transports to the middle brain. One day after injection, the CA in the brain parenchyma and CSF was washed out (Figure 2A).

Then, metabolism of CA was evaluated in vivo with mice infected by rAAV2. CA was administrated intrathecally with 200nl volume in the left brain and allowed infection for 21 days. In vivo T1W images were collected on two mice at different time points using a 3T MRI scanner, as shown in Figure 2B and C. After injection, CA was gradually diffused to brain parenchyma and ventricles. Blue circles in Figure 2B and C indicate hyperintense regions that were gradually appeared with time on T1W images. The co-localization of hyperintense regions on MRI and fluorescence images are shown in Figure S2, indicating that signal increments on T1W MRI images are due to the uptake of Gd-EOB-DTPA by OATP1A1. One day after injection, CA in those brain regions without OATP1A1 expression was washed out, while the contrasts in those brain regions with OATP1A1 expression retained (Figure 2B and C). One week after injection, CA in the whole brain was washed out (Figure 2B). MR signals were averaged within thalamus. Mean T1W signal intensities in left thalamus (OATP1A1 expression site: blue circles in Figure 2B and C) and right thalamus (control site: red circles in Figure 2B and C) are compared. The largest differences between MRI signals in OATP1A1-expressing regions (left) and control regions (right) happened one day after CA injection. The same phenomenon was also observed at 14T (Figure 2D). After intrathecal injection of CA, signals in Lateral posterior thalamic nucleus (LP) and

Secondary somatosensory cortex (S2) areas of the left hemisphere showed high signals. Signal gradually increased and returned to the baseline one week later. The largest contrasts between OATP1A1-expressing regions (left) and control regions (right) happened one day after CA injection, which is consistent with 3T results.

A
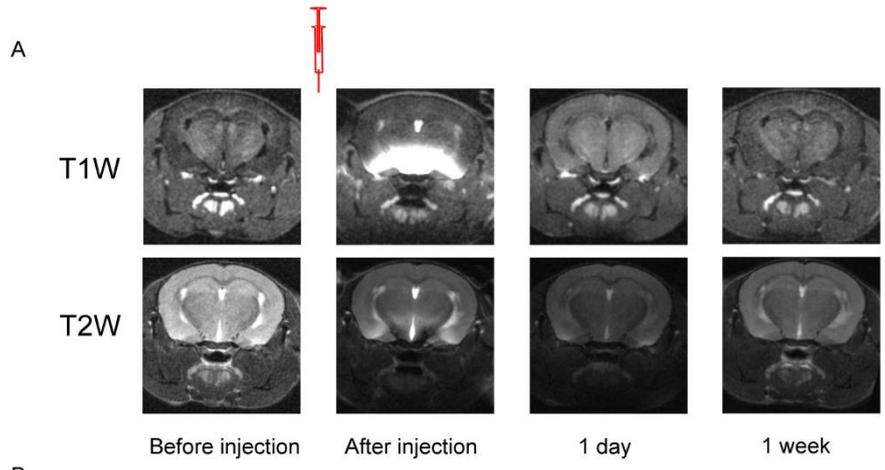

B
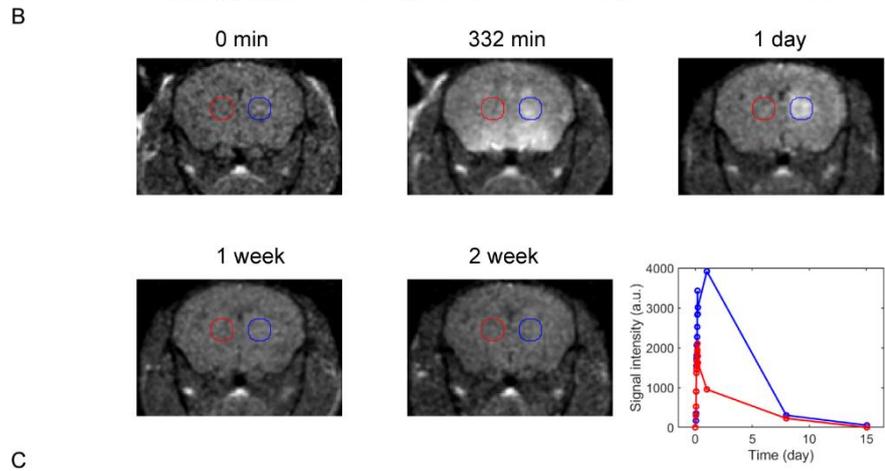

C
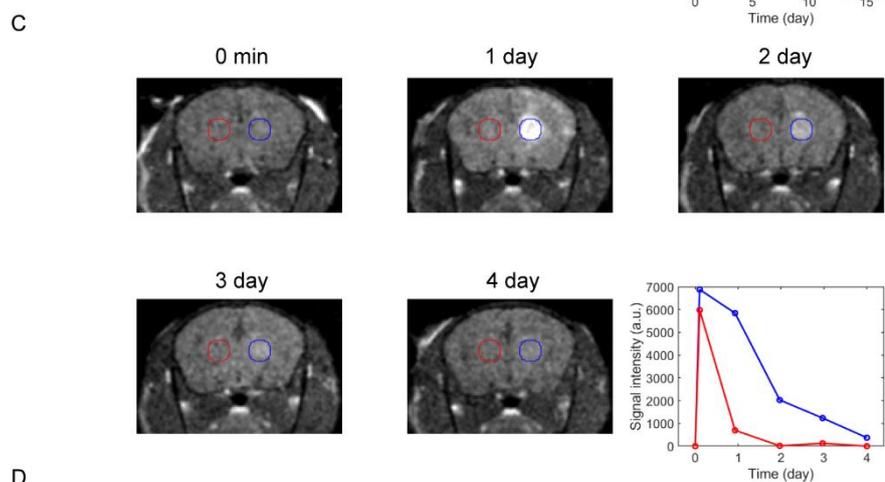

D
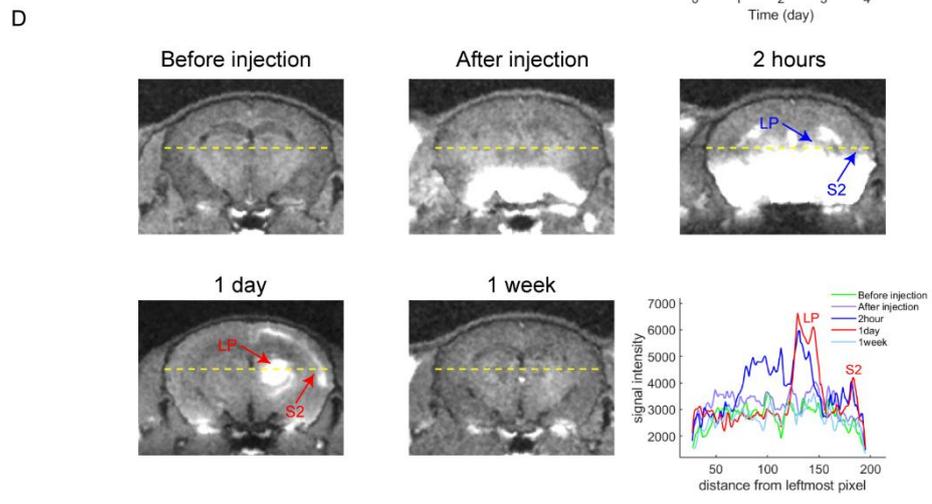

Figure 2 Diffusion and metabolism of Gd-EOB-DTPA in the brain. Data were collected from four mice (mouse-1, mouse-2, mouse-3 and mouse-4). (A) T1W and T2W images were collected at 14T from mouse-1. All images were registered to each other. CA diffuses from the lower part of the skull to upper portion of the brain, and washed out one week after CA injection. (B) T1W images collected on a 3T scanner at different time points from mouse-2, where 0 min represents the time point right before injection of Gd-EOB-DTPA. ROIs indicating the OATP1A1-expression-site (blue) and control-site (red) are also overlapped on the images. Dynamic curves of signal intensities are plot and compared. (C) T1W images and dynamic curves of signal intensities from mouse-3 are shown. In this case, data were collected at different time points on a 3T scanner within one week. (D) Similar results were found from data collected from mouse-4 at 14T. Signals (indicated by the dashed yellow lines) collected at different time points (before injection, after injection, 2 hours, 1 day and 1 week after injection) showed similar results compared with 3T. Two ROIs (LP and S2) were selected for comparison and marked by arrows.

**Virus infection and CA injection have no visual damage of brain parenchyma integrity on T2W**

To evaluate the integrity of brain parenchyma after virus infection and CA injection, high resolution T1W and T2W images were collected at 14T. Figure 3A shows T1W and T2W images collected one day after CA injection. The virus injection site is indicated by yellow arrows on T1W and T2W images. Red arrows indicate brain regions projected to PO. T2W images collected at the same time didn't show any abnormalities in the brain parenchyma, except the needle track damage created during the stereotactic injection. Some regions with enriched CA did show low signals, which might be caused by T2 relaxation effect of CA. However, as CA completely washed out one week after injection, hypointense signals disappeared on T2W, and both T1W and T2W images return to the baseline (Figure 3B).

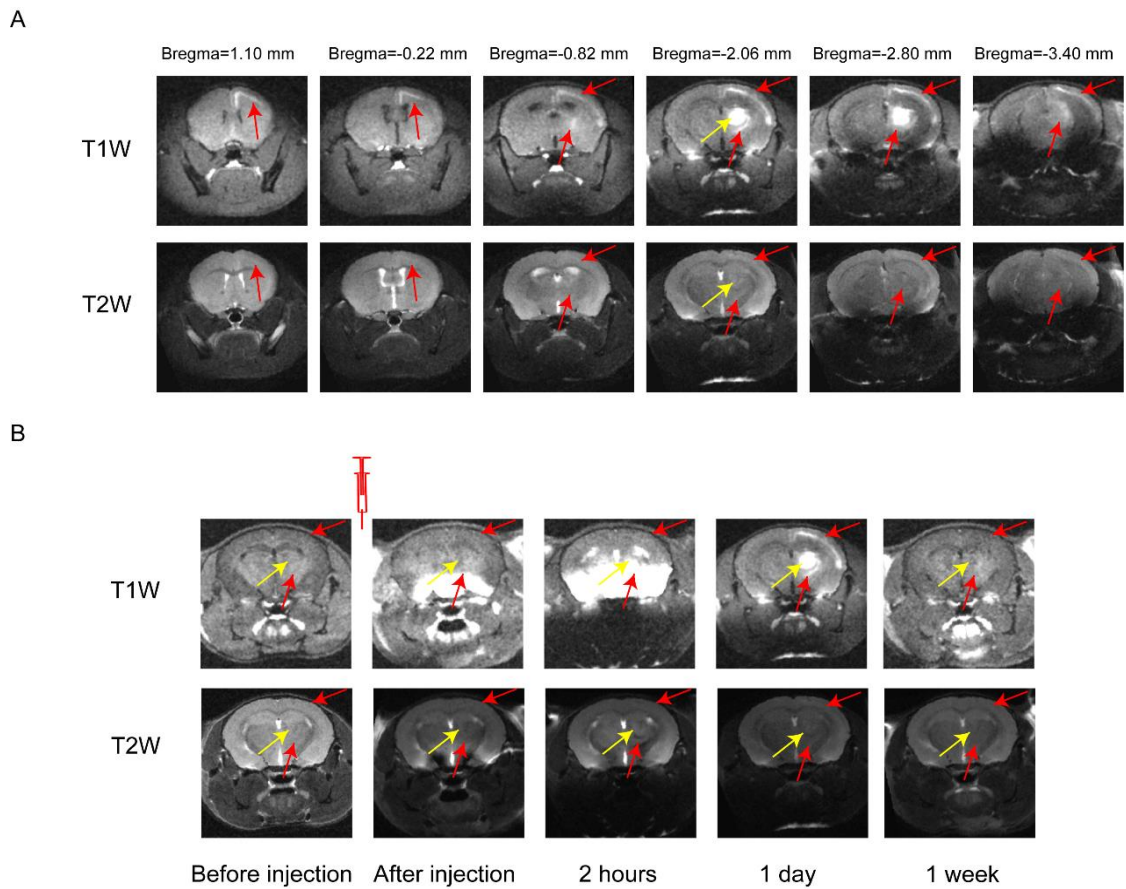

Figure 3 Evaluation of brain parenchyma integrity using high-resolution T1W and T2W images collected at 14T. All images are registered. (A) T1W and T2W images collected one day after intrathecal injection of CA. Images of different slices are shown. Red arrows indicate brain areas projected to PO, and yellow arrows indicate the injection

site. (B) T1W and T2W images collected at 5 key time points show dynamic changes of the CA in brain parenchyma. Tissue abnormalities were evaluated using T2W images. Red arrows indicate brain regions projected to PO, and yellow arrows indicate the injection site.

**MRI and fluorescence imaging show comparable results of neuronal labeling**

To demonstrate the accuracy of MRI based neuronal labeling, MRI results were compared with the fluorescent results. Two mice were injected with 300nl rAAV2-retro-oatp1a1-P2A-EGFP and rAAV2-retro-EGFP viruses into the PO areas, respectively. After waiting for twenty-one days, 7ul Gd-EOB-DTPA was administrated intrathecally. T1W images collected at 14T are shown in Figure 4A. Yellow arrows in Figure 4A indicate the uptake of Gd-EOB-DTPA occurs only in the OATP1A1-expressing regions. T1W images were also compared between a EGFP virus infected mouse and a normal mouse (Figure S3), and no hyperintense contrasts in the cortex were found in either case. Comparison between MRI and fluorescent results are shown in Figure 4B. It shows that neuronal labeling on MRI coincides with the fluorescent results, which is considered as the golden standard for neuronal labeling. To further demonstrate the accuracy of MRI based neuronal labeling, we compared two methods using a single mouse. We injected 200nl rAAV2-retro-EGFP into left PO (1.30, -2.03, -3.15) and 200nl rAAV2- retro-oatp1a1-P2A-EGFP into right PO (-1.30, -2.03, -3.15) (Figure 4C). Both fluorescent and MRI images show projection of neurons from cortex to PO in both hemisphere (Figure 4D).

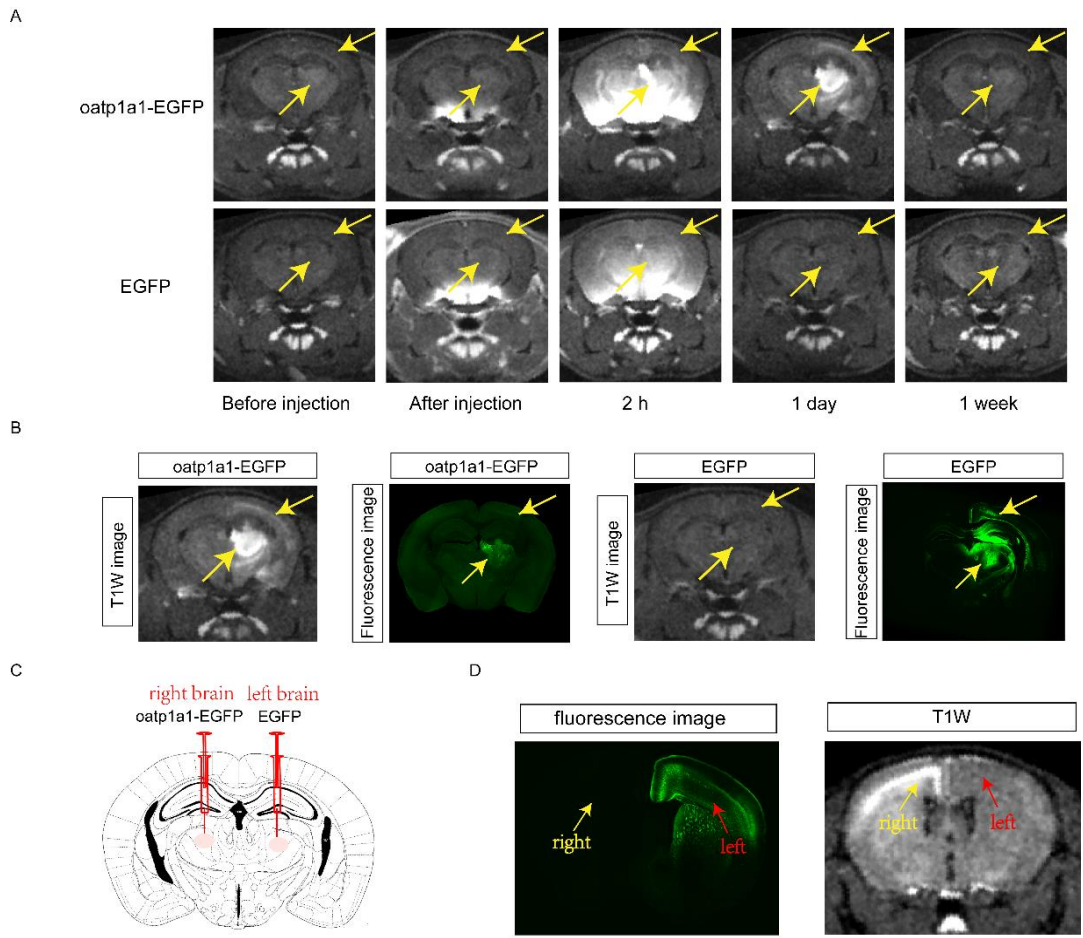

Figure 4 Comparisons between MRI and fluorescent results. (A) T1W images collected at 14T from two mice infected with two viruses (upper: rAAV2-retro-oatp1a1-P2A-EGFP; lower: rAAV2-retro-EGFP), respectively. Images were collected at five critical time points. (B) MRI images are compared with fluorescent images. All images were collected from the same mice used in (A). Yellow arrows indicate the injection sites and cortical regions with hyperintense signals. (C) Schematic diagram of virus injection, 200nl of rAAV2-retro-EGFP was injected into the left brain, and 200nl of rAAV2-retro-oatp1a1-P2A-EGFP was injected into the right brain. (D) MRI images are compared with fluorescent images on a single mouse. Yellow and red arrows indicate the injection

sites for the two viruses (left: rAAV2-retro-oatp1a1-P2A-EGFP; right: rAAV2-retro-EGFP), respectively.

**Imaging of labeled neurons in the whole brain using in vivo MRI**

Twenty-one days after virus injection (200 nL volume), experimental mice were intrathecally administered with Gd-EOB-DTPA. All MRI images were collected one day after CA injection. T1W images were registered to the TMBTA-Brain-Template and TMBTA-Brain-Atlas(Hagler et al., 2019). As shown in Figure 5A, SSp and SSs regions show high signals, which is consistent with previous reports(Huo et al., 2020; Oh et al., 2014) indicating that these regions contain neurons project to PO. Hyperintense signals were also found in RT (reticular nucleus), MOs (Secondary motor area), MOp (Primary motor area) and VIS(visual area)(Huo et al., 2020; Martinez-Garcia et al., 2020; Oh et al., 2014), all of which have neurons projected to PO. Signals in these brain regions were compared to contralateral regions (Figure 5B), and significant differences were found (with p-values << 0.001). No significant differences were found in EPv, indicating no projections from this region to PO (Figure 5B).

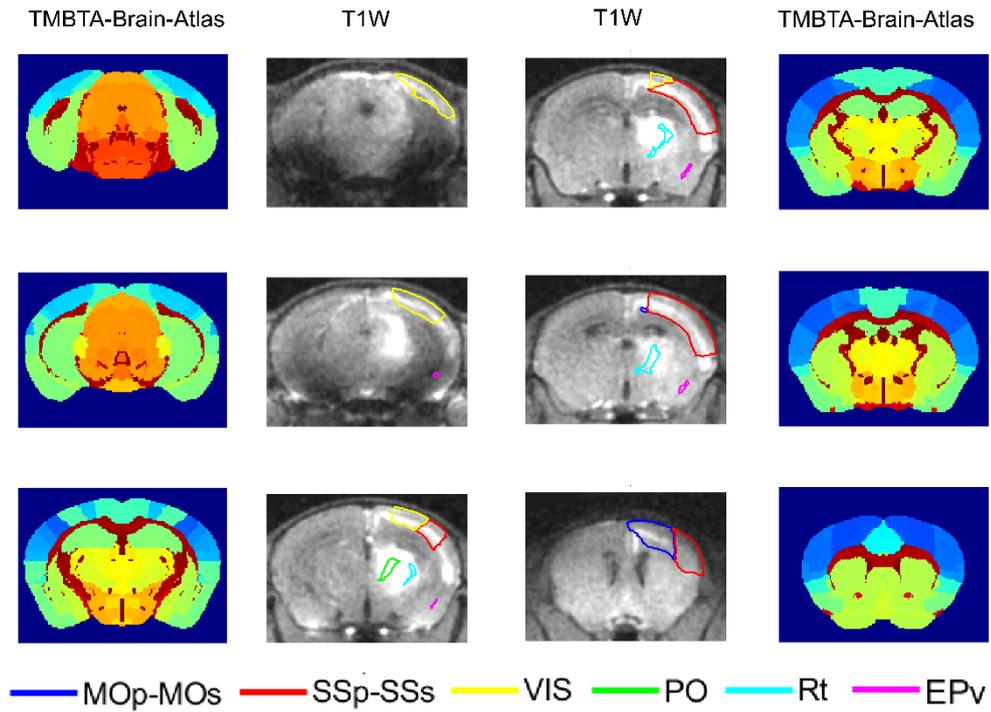

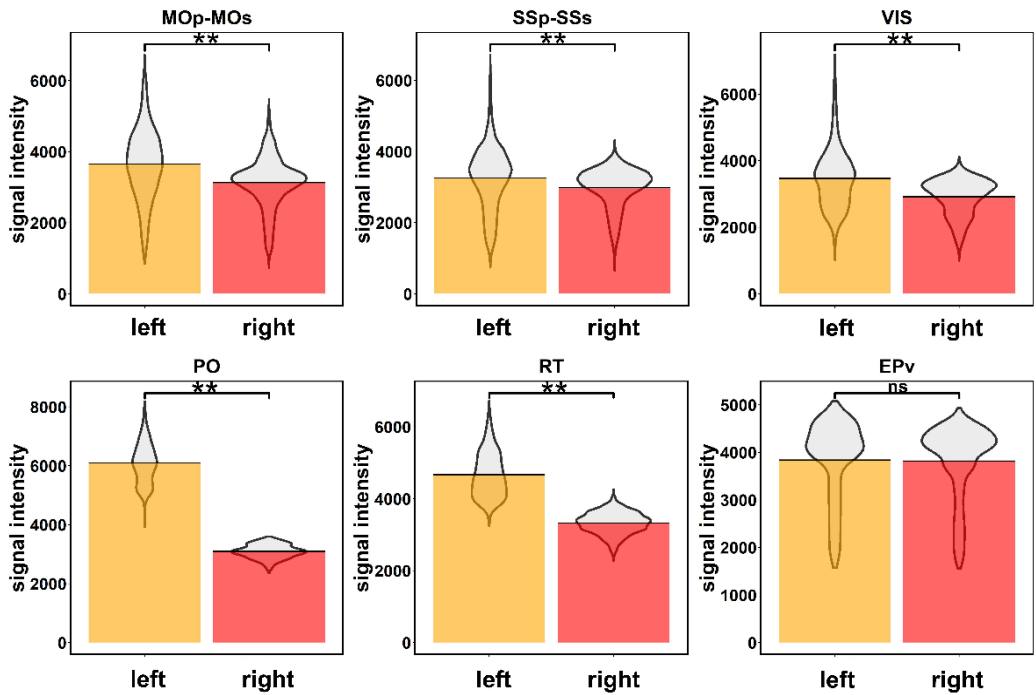

Figure 5 Regional comparison of T1W signals. T1W images were registered to the TMBTA-Brain-Template and ROIs were obtained from the TMBTA-Brain-Atlas. (A) T1W images were registered to the TMBTA-Brain-Template. ROIs (SSp, SSs, MOp, MOs, RT, VIS, PO) from the TMBTA-Brain-Atlas were overlapped on the images and marked with different colors. (B) Comparisons of T1W signal intensities between the left and contralateral regions of the 6 ROIs. P-values of the comparison are shown above each plot. (P<<0.001: **; 0.001<P<0.01: *; P>0.01: ns)

**DCE based analysis allow quantification of neuronal labeling**

Tofts model was used to fit the Gd concentration curves to obtain quantitative parameters $K$ and $v_t$, as shown in Figure 6. Regions that contain labeled neurons show significantly higher $v_t$ values compared to those regions that don't contain labeled neurons, indicating that parameter $v_t$ is a useful biomarker for neuronal labeling.

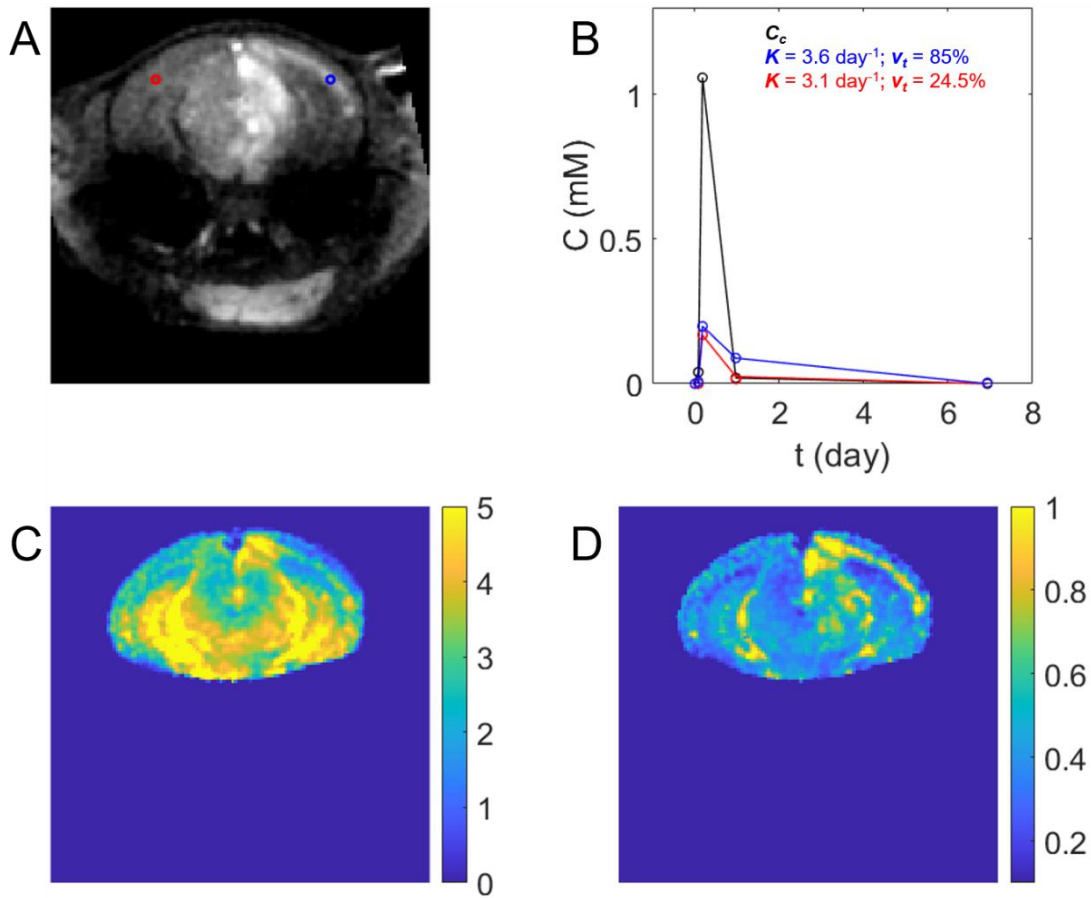

Figure 6 Quantitative evaluation of labeled neurons using Tofts model. (A) T1W image

acquired at 1397 min. The fitting curves of two pixels, located in the regions with (blue) and without (red) projections, are shown in (B). The curve of Gd concentration in CSF ($C_c$) is marked by black color. Fitting parameters $K$ and $v_t$ are shown in (C) and (D), respectively.

**Discussion**

Fluorescent indicators provide an optical approach to label and image neurons. Although advanced techniques, such as multi-photon-microscopy, have the ability to image deep in light-scattering tissue, its observation depth is still limited to around 1 mm (Lecoq et al., 2019). Due to this intrinsic penetration limits, optical techniques in most cases require sacrifice of the animal and histological processing of the ex vivo tissue. As a result, a large sample size is usually required, and the brain slicing process might also cause damage of the brain tissue. Especially in the case of achieving whole brain neuronal mapping, special high-throughput techniques, such as fluorescence micro-optical sectioning tomography (fMOST) (Gong et al., 2013; Sun et al., 2019), volumetric imaging with synchronized on-the-fly-scan and readout (VISoR) (Wang et al., 2019; Xu et al., 2021) and transparent brain technology (Cai et al., 2019; Chung et al., 2013), are usually required.

Mammalian brain is a complex organ, and our understanding of it is still limited. Brain contains tens of billions of neurons and they form hundreds of thousands of neural circuits to perform functions. Although single neuron studies have greatly promoted the progress of neuroscience, people have long realized that these findings must be placed in the context of natural physiological networks of neural circuits. In order to study the dynamics of neural circuits rather than just studying the impact of networks on single-neuron behavior, developments of new techniques that allow simultaneous in vivo labeling and quantification of large amounts of neurons are necessary. Calcium imaging (de Melo Reis et al., 2020) with fluorescent indicators provides an optical approach to label neurons and measure neuronal activity in vivo. However, it requires complex surgery to open the skull for observation, which might induce inflammatory reactions (Holtmaat et al., 2009; Rose, 2012), and whole brain mapping of labeled neurons using calcium imaging is still challenging.

In this study, we propose a new in vivo strategy for neuronal labeling, imaging and quantification. As a demonstration, we use neurotropic virus to deliver a reporter gene oatp1a1 into targeted neural circuits, and show its ability to label and image neurons in vivo, which provides an efficient way to obtain whole brain mapping of labeled neurons in targeted neural circuits. Our results show that oatp1a1 can be used as an efficient reporter gene, which could be potentially useful for structural and functional neuroscience studies when combined with in vivo MRI. Although the imaging resolution is limited compared to optical techniques, MRI provides a unique opportunity for in vivo neuronal observation. This is crucial for observing the dynamic changes of neurons in individual animals, and can be an effective supplement to optical methods. As the field strength goes higher and higher, resolution of MRI has been pushed to new limits, which make it possible for MRI to achieve cellular level resolution in the foreseeable future. Importantly, the method proposed in this study can provide quantitative analysis of neuronal labeling.

In conclusion, we provide in this study a new method for neuronal labeling. When combined with MRI, it allows in vivo neuronal observations and is expected to have various application prospects in studying basic neuroscience and chronic nervous system diseases.

## Methods

### Animal preparation

Animals involved in this study were treated in accordance with protocols approved by the Animal Ethics Committee at University of Science and Technology of China (SYXK2017-005). Male C57BL/6J mice (6–8 weeks old) were purchased from SPF (Beijing) Biotechnology Co., Ltd. All animals were reared in a 12 hr/12 hr light-dark cycle room, at a suitable temperature, and food and water are freely available.

### Virus construction

The oatp1a1 gene sequence was acquired from NCBI nucleotide database (reference sequence NM_013797.5). Recombinant plasmid of the virus (rAAV-EF1a-oatp1a1-P2A-EGFP-WPRE-pA and rAAV-EF1a-EGFP-WPRE-hGH-pA) was designed in-house and packaged virus was purchased commercially from BrainVTA (Wuhan, China). The virus rAAV2-retro provides retrograde (from axon to soma) infection of neurons. For neurons infected by the rAAV-EF1a-oatp1a1-P2A-EGFP-WPRE-pA virus, both EGFP and OATP1A1 proteins are expressed. For neurons infected by the rAAV-EF1a-EGFP-WPRE-hGH-pA virus, only EGFP proteins are expressed. The virus titer was detected using SYBR Green qPCR method. The titer of both viruses were 6.09E+12vg/ml, respectively.

**Virus injection**

Before surgery, experiment mice were fixed in the stereotaxic device (RWD, Shenzhen, China) and ketamine was administrated to relieve the pain. During operation, the mice were anesthetized with isoflurane gas using an anesthesia machine (RWD, Shenzhen, China). Mouse body temperature was maintained by a heating pad. Glass microelectrode was connected to the micro syringe, and virus was injected into the PO (posterior thalamic nuclear group) area: 3.15 mm dorso-ventral (DV), 2.03mm anterior-posterior (AP), 1.30 mm medio-lateral (ML). The total volume of injected virus was controlled between 200-300nl, and the injection speed was set to 30nl/min. After injection, the glass microelectrode was kept at the injection site for 5 minutes before removing. Skins around the operation site were sutured and sterilized after surgery. At the end of the experiment, gas anesthesia was stopped and experiment mice were put back into the cage.

**Fluorescence imaging and immunofluorescence**

Mice infected with the virus more than 21 days were used for ex vivo experiments.

The mice were anesthetized with isoflurane, and the heart was perfused with PBS and 4% PFA. The brain was extracted and soaked in 4% PFA overnight. After that, the brain was dehydrated with sucrose solution. First, 20% sucrose solution was used for dehydration and sedimentation, and then 30% sucrose solution was used for dehydration and sedimentation. The dehydrated brain was sliced coronally using freezing microtome (Leica, CM1950) and OCT (embed brain tissue). For fluorescence imaging, the slice thickness is 40um, and for immunofluorescence the slice thickness is 30um. The brain slices were stored in antifreeze solution (30% ethylene glycol, 20% glycerol, 50%PBS).

For fluorescence imaging, brain slices were washed with PBS (3 times, 5 minutes each time) in 24-well plates and then flatten on a glass slide. Finally, the brain slices were mounted with anti-fluorescence quencher (CC/MountTM SIGMA) and a cover glass. For immunofluorescence, the sections were washed with PBS (3 times, 5 minutes each time) in a 24-well plate and then blocked using blocking buffer for immunostaining (Proteintech catalog number PR30008, room temperature, shaker blocked, 1h). The sections were incubated with the primary antibody (Anti-Slco1a1 Polyclonal Antibody, 1:300, Rabbit, Bioss) overnight at 4 °C. After that, sections were wahsed with PBST (3 times, 5minutes each time) and incubated with the secondary antibody (CL594 – conjugated Goat Anti-Rabbit IgG(H+L), 1:500, Goat, Proteintech, room temperature, 1h, protect from light). After washing the sections (PBST, 3 times, 5minutes each time), DNA were stained with DAPI (meilunbio MA0128) in a dark environment for 10 minutes. The sections were then washed with PBS (2 times,

5minutes each time) and anti-fluorescence quencher was added dropwise to mount the sections. Nikon Ti2-E confocal microscopes, FV-3000 confocal microscope and SLIDEVIEW VS200 were used for fluorescence imaging of the sections.

**Plasmid construction**

pcDNA3.1 was a vector skeleton with mCherry. The oatp1a1 gene was amplified from the viral plasmid (rAAV-EF1a-oatp1a1-P2A-EGFP-WPRE-pA). And the full-length of oatp1a1 gene was cloned into the vector skeleton by homologous recombination. First, forward and reverse primers were ordered from GENERAL BIOL company according to the vector map (oatp1a1-mCherry: forward: GAAAACTAAGCTGTGAGAATTCTGCAGATATCCAGCAC; reverse: TCTCTGTTTCTTCCATGGATCCCTTGTACAGCTCGTCC. oatp1a1: forward: ATGGAAGAAACAGAGAAAAGG; reverse: TCACAGCTTAGTTTTCAGTTCTC). The conditions of 50ul PCR system were set as follows: 98°C for 5min; 35 cycles of 98°C for 30s, 58°C for 30s and 72°C for 2min30s; and 72°C for 20min. After PCR, we do the DNA gel electrophoresis and gel recovery to recover the target fragment. Second, ClonExpress MultiS One Step Cloning Kit (Vazyme) was used for homologous recombination (5X CE Multis Buffer=2ul, Exnase Multis=1ul, PCR products of oatp1a1 fragment, pcDNA3.1- oatp1a1-mCherry vector backbone with homology arms and ddH$_2$O=7ul). After centrifugation, put the reaction system in PCR（37°C for 30min）. Third, transfer the constructed plasmid (10ul) into DH5α competent cells. Than DH5α was incubated on ice for 20-25min for plasmid enrichment and hit by heat 1min30s on 42°C water bath. After that, 1ml of LB medium was added to the DH5α competent cells and all of these were put in 37°C shaker for 1h. After centrifugation, spread the bacterial solution in ampicillin-resistant LB medium (1:1000) and incubate overnight at 37°C incubator. Pick and cultivate monoclonal bacteria and send them to the company (GENERAL BIOL) for sequencing. After sequence alignment, the plasmid is extracted for cell experiment (Plasmid extraction kit AXYGEN).

**Cell experiment**

SH-SY5Y cells were transfected with 2 μg cDNA of Oatp1a1-mCherry using jetPRIME transfection reagent (Polyplus) for 24h and plated onto poly-L-lysine-coated culture dishes with glass bottom. Images were taken at 594 nm for mCherry using a confocal system consisting of an Eclipse Ti inverted microscope (Nikon), a CSU-X1 Spinning Disk unit (Yokogawa), a DU-897U EMCCD camera (Andor), a laser-controlling module (Andor) and iQ3 imaging software (Andor) with a 100× oil immersion lens.

**In vivo MRI**

3T (GE Discovery 750W) and 14T (Bruker AVANCE NEO 600 WB) scanners were used for MRI. The mice were first induced anesthesia with 1%-1.5% isoflurane and 0.3-0.5L/min air, and then maintained anesthesia with 1%-1.5% isoflurane and 0.3-0.4L/min air during imaging. A commercially available volume coil (Medcoil, Suzhou,

China) was used for 3T imaging. For 3T imaging, 2D T2 Fast Spin Echo (T2-FSE) sequence (TR = 6574.0 ms; TE = 102.0 ms; pixel size = 187.5×187.5 µm$^2$; slice thickness = 0.9 mm; scan time = 2 min) was used to acquire T2W images. A 2D T1 Spin Echo (T1-SE) sequence (TR = 600ms; TE = 14 ms; Freq. pixel size = 156.3×156.3 µm$^2$; slice thickness = 0.6 mm; scan time = 14 min) was used to acquire T1W images. For 14T imaging, standard Bruker 2D T1-RARE-Inv-Rec sequence (TI = 1.05 s; TE = 5 ms; TR = 3.8 s; pixel size = 78.1×78.1 µm$^2$; slice thickness = 0.5 mm; scan time = 12 min) was used to acquire T1W images. A 2D T2-TurboRARE sequence (TE = 2.8 ms; TR = 2.8 s; pixel size = 58.6×58.6 µm$^2$; slice thickness = 0.5 mm; scan time = 9 min) was used to acquire T2W images.

All MRI images were collected before and after CA injection. Gd-EOB-DTPA (5ul-8ul, Bayer) was injected into the spinal cord of mice by intrathecal injection. After intrathecal injection, keep the needle (0.5ml, 29G x ½, KRUUSE) at the injection position for 5 minutes. After CA injection, T1W and T2W images were collected at different time points.

**Data analyses**

MATLAB (R2021a) was used for the analysis of MRI images. T1W and T2W images were registered using FLIRT (Oxford, UK) in FSL (Jenkinson et al., 2002; Jenkinson and Smith, 2001). Because FSL is designed for human MRI image processing, resolution of all images collected on mice were increased by 10 times before registration. Region of interest (ROI) was drawn on the MRI images, and statistics (mean, median and standard variation) were calculated within each ROI. MRI images were also matched onto a mouse brain atlas (a public database of mouse brain template TMBTA) using affine registration. Targeted ROIs include PO, RT, SSp, SSs, MOs, Mop, VIS and EPv. For comparison purpose, slices of fluorescent images with similar anatomical structures were used to match MRI results. In Figure 4, the p-values were calculated with t-test in R.

**Quantitative analyses**

T1_RARE_Inv_Rec sequence was used to acquire T1W images on a 14 T scanner (Bruker AVANCE NEO 600WB). The approximate MR signal intensity $S$ is given by

$$S = S_0 \cdot [1 - 2\exp(-R_1 \cdot TI) + \exp(-R_1 \cdot TR)] \cdot \exp(-R_2 \cdot TE) \quad (1)$$

where $S_0$ is a scaling factor which is proportional to spin density. $TI = 1.05$ s and $TR = 3.8$ s represent inversion recovery interval and repetition time, respectively. $R_1$ and $R_2$ are longitudinal and transverse relaxation rate constant, respectively. Dividing signals of the $n^{th}$ time point ($S_n$) by signals before injection, $S_0$ and $R_2$ terms are automatically eliminated

$$\frac{S_n}{S_1} = \frac{1 - 2\exp[-(R_{10} + r_1 C) \cdot TI] + \exp[-(R_{10} + r_1 C) \cdot TR]}{1 - 2\exp(-R_{10} \cdot TI) + \exp(-R_{10} \cdot TR)} \quad (2)$$

It should be noted that accurate determination of Gd concentration ($C$) requires $R_1$ (= $1/T_1$) mapping. Here for demonstration purpose, we assume $R_{10} = 1$ s$^{-1}$ and use $r_1 = 6.9$ s$^{-1}$mM$^{-1}$ for Gd-EOB-DTPA, $C$ can be determined using Eq. (2).

The flow of Gd-EOB-DTPA from CSF into tissue space is determined by the standard Tofts model (Tofts and Kermode, 1991):

$$v_t \frac{dC_t(t)}{dt} = K \cdot [C_c(t) - C_t(t)] \qquad (3)$$

where $C_t$ and $C_c$ represent Gd concentration in tissue and CSF, respectively. $K$ is the transfer constant, and $v_t$ is the fractional volume of tissues that uptake Gd-EOB-DTPA. The solution of Eq. (3) is given by:

$$C_t(t) = K \int_0^t C_c(\tau) \cdot \exp\left[-\frac{K}{v_t}(t-\tau)\right] d\tau \qquad (4)$$

Images were acquired at five time points before (t = 0 min) and after (t = 114, 258, 1397 and 9996 min, respectively) Gd injection. $C_c$(t) is averaged over $C$ values within the 4th ventricle. Quantitative parameters $K$ and $v_t$ can be determined by fitting the experimental curves to Eq. (4).

**Acknowledgements**

We thank the staff (Ke Gong, Yusong Wang) at the Instruments Center for Physical Science, University of Science and Technology of China, for helping with 14T MRI experiments setup. This work was partially supported by the National Natural Science Foundation of China (22077116), the Collaborative Innovation Program of Hefei Science Center, CAS (2020HSC-CIP010), the USTC Research Funds of the Double First-Class Initiative (YD9110002011), and the Fundamental Research Funds for the Central Universities (WK9110000119).